\documentclass[%
 reprint,
 amsmath,amssymb,
 aps,
]{revtex4-2}
\usepackage{amsmath}
\usepackage{mathtools}
\usepackage[T1]{fontenc}
\usepackage[utf8]{inputenc}
\usepackage[brazil]{babel}
\usepackage{upgreek}
\usepackage{mathtools}
\DeclareUnicodeCharacter{2212}{-}
\usepackage{graphicx}
\usepackage{dcolumn}
\usepackage{bm}
\usepackage{float}
\usepackage{caption}
\usepackage{subcaption}

\begin{document}

\preprint{APS/123-QED}

\title{A Theoretical Analysis of Superconducting Pairing in Correlated Metallic Systems}%

\author{Koushik Mandal }
\altaffiliation[]{koushikmandal@bose.res.in; iamkoushik424@gmail.com \\
 }
\author{Ranjan Chaudhury}%
 \email{ranjan@bose.res.in}
\affiliation{%
 S N Bose National Centre for Basic Sciences \\
 Block- JD; Sector-III, Salt Lake, Kolkata, India\\
   \textbackslash\textbackslash
}%




\date{\today}
\begin{abstract}
We have introduced a Coulomb correlated normal state  to study the superconducting pairing within a Fermi liquid(FL)-like background in a 3-dimensional system. The role of the Coulomb correlation has been actively incorporated by means of the Gutzwiller projection scheme, in the presence of phonon mediated attractive electron-electron(AEE) interaction throughout the entire regime of its applicability. We variationally modulate the Coulomb correlation strength from the weak to the strong regime in our calculational procedure. The main highlight of our results is the appearance of a `2-gap-like' structure in the superconducting phase, arising out of the two-body interacting matrix elements, calculated in the presence of the active Coulomb correlation. We have made use of the `two-square well model' potential to evaluate the superconducting pairing gaps at zero temperature. Our Calculational scheme has been applied to a simple cubic lattice system for a physical realization. All the consequences are elaborated and discussed.
 
\end{abstract}

\maketitle
\section{\label{sec:level1}Introduction}
 Superconductivity is one of the most mysterious physical phenomena discovered in the last century. Even today, it is a challenge to prepare a stable room temperature superconducting (RTSC) system. Meanwhile, in the last year, Dias et.al. had found a superconducting state at a temperature of 287.8K in a pressure driven carbonaceous sulfur hydride system\cite{dias}. The other high-temperature superconductors (HTSC), with a critical temperature($T_c$) close to the ambient temperature, were mainly found in the pressure induced superhydride systems ($e.g.$ $ LH_{10}, YH_{x}$)\cite{shylin,memley,duan, hamlin,errea}. Their promisingly high $T_c$, is due to the electron-electron attractive coupling mediated by the high frequency phonon which is usually generated in the materials having elements with very low atomic masses\cite{ashcroft1,ashcroft2,bonev, mcmahon,mcminis,parks,allen}. \\
The original microscopic theory of Bardeen-Cooper-Schrieffer (BCS) had proposed pairing from phonon mediated electron-electron attraction
 \cite{schriefferp,schriefferb,tinkham}.\\
The role of the repulsive Coulomb correlation in the pairing theory was analyzed within the BCS framework much later\cite{daemen}.  However, our aim here is to examine the superconducting pair formation in the s-wave channel from a Coulomb correlated parental normal phase itself. This will bring out up to what strength of the Coulomb correlation in the normalphase, the system can sustain  superconductivity. \\
In the recent past, R. E. Zillich et.al. have reported  the BCS pair formation in a correlated low-density Fermi gas using a square-well potential as well as Lennard-Jones (LJ) potential\cite{zillich}. They defined a  correlated variational state to take into account the Coulomb correlation, as this scheme was previously implemented to study correlated superfluid $He^{3}$\cite{paulick,sorella,krot}.\\ 
In literature, the competition between the phonon driven AEE interaction and the  Coulomb correlation was reported earlier in Holstein-Hubbard (HH) model  for different lattice systems\cite{Holstein, Hubbard,Alexandrov,costa}.\\
The Gutzwiller approximated projection scheme as was earlier proposed by M. C. Gutzwiller to study the ferromagnetism in the itinerant d-band electrons in rare-earth oxide materials, is a projected Slater determinant that minimizes the real space configuration with the prevention of doubly occupied sites. \cite{gutzwiller1963,gutzwiller1964,gutzwiller1965}. The approximated Gutzwiller wave function has been extensively studied in different lattice dimensions and therefore also been implemented to study superconductivity using the Hubbard model with an attractive interaction
\cite{vollahardt1988,gebhard1990,bunemann1998,bunemann2005,bunemann2016}.\\
Meanwhile, we have studied the inter-playing characteristics of the phonon mediated AEE interaction and the Coulomb interaction in a superconductor with both weakly and strongly correlated situations\cite{km}. We had proposed two distinct ways to construct the correlated variational state and one of which was elaborated in our previous work\cite{km}. In ref.\cite{km}, the  correlated pairing state was configured with the BCS paring operator acting on the Fermi sea (FS) ground state to form a paired state and then the Gutzwiller operator acting on the paired state. In other words, the Coulomb correlation involves the paired states.\\
Subsequently, we put forward another way to construct the correlated BCS (CBCS) pairing state. The variational pairing state is set up starting with a correlation induced FS ground state $i.e.$ correlated Fermi sea (CFS)  ground state. The Gutzwiller partial projection operator with a variational parameter ($\alpha$) takes care of the Coulomb correlation by reducing the amplitude for the double occupancy on a site. The BCS pairing operator is therefore applied on this CFS to form the CBCS state.\\
\begin{figure}
\centering
\includegraphics[scale = 0.65]{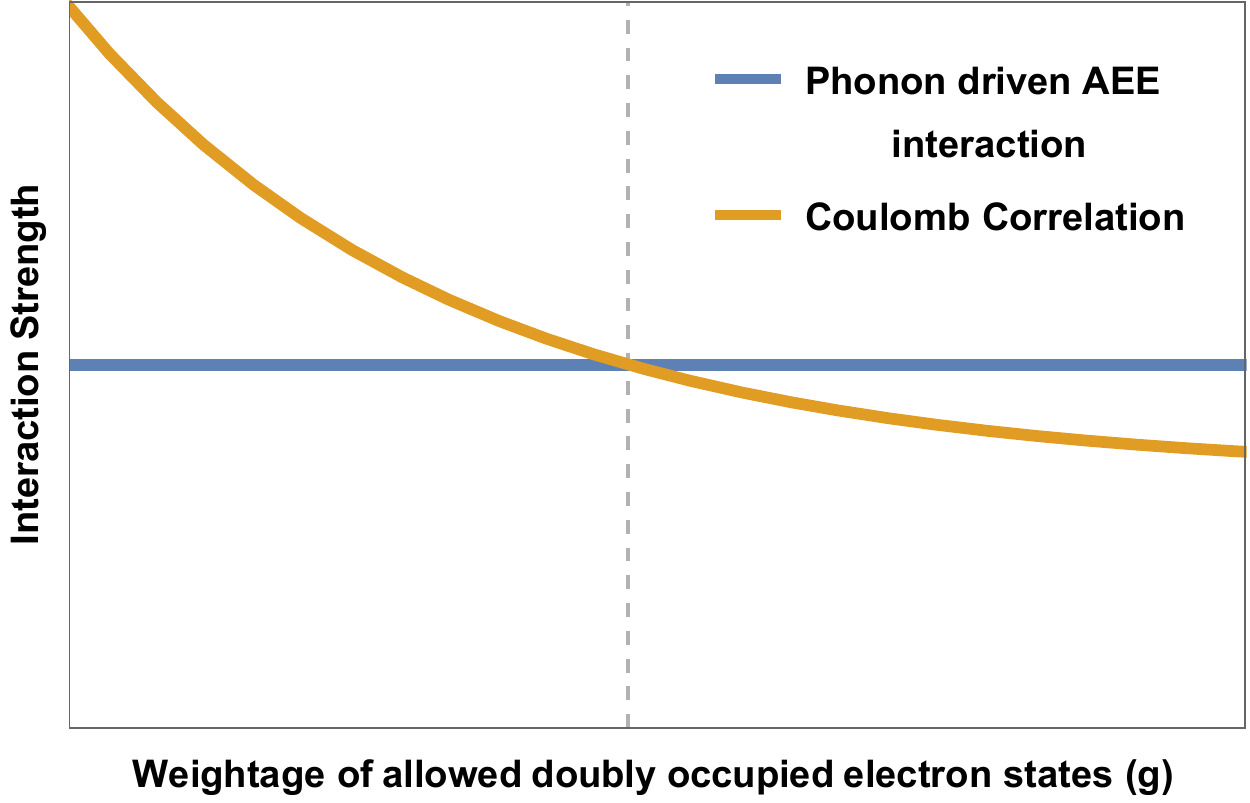}
  \caption{Schematic phase diagram: Attractive phonon driven electron-electron interaction and Coulomb interaction with the weightage of the allowed doubly occupied sites\cite{lado}.}
  \label{fig:boat1}
\end{figure}
 Since, our primary goal of defining the variational CBCS state is done, we would like to 
apply it to a real electron system to find the sustainability of isotropic superconducting pairing in the presence of Coulomb correlation. The expected qualitative behaviour of Coulomb correlation and phonon induced AEE interaction with the weightage of the allowed doubly occupied site is schematically shown in fig:1\cite{lado}. The phonon driven AEE interaction is independent of the weightage whereas the Coulomb interaction decays as a function of the same weightage. Hence, there is a crossover at a certain weightage value, beyond which attractive phonon driven AEE interaction dominates over the repulsive Coulomb interaction. This weightage can now be controlled by the Gutzwiller parameter which reduces the amplitudes for the double occupancy on a site. A critical value of this parameter will decide the phase separation between the superconducting phase and the correlated normal state. This transition point implies that there must be an upper cut-off value in terms of the Coulomb correlation parameter ($\alpha$) beyond which SC-state is no longer possible.\\
This manuscript is configured as follows: In \textit{Sec-2}, we formulate the variational CBCS state and the relevant model Hamiltonian to study the superconducting pairing instability in correlated fermions on a simple cubic lattice. Based on this defined variational CBCS state, we evaluate the total ground state energy and the superconducting pairing gap-equations. All the analytical calculations and results are elaborated in \textit{Sec-3}. Finally, we present the conclusions of our results and parallely analyse the outcomes in combination with those of our previously reported work in \textit{Sec-4}. At the end, \textit{Appendix-A}, \textit{Appendix-B}, and \textit{Appendix-C} are included to discuss the normalization of the CBCS state, the evaluation of the total ground state energy, and the equations for the superconducting pairing gaps with a few major calculational steps shown.
\section{\label{sec:II}Formulation of the problem}
\subsection{\label{sec:level2} Correlated Fermi sea ground state}
 The Gutzwiller wave function in detail, is a  variational wave function for many
particles system that describes interacting particles on a rigid lattice systems  with broken ground state symmetry ($e.g.$ N$\'{e}$el state for the anti-ferromagnetic system). By considering this constraint  as well as  the approach of the introduction of the Coulomb correlation, we need to define the variational CBCS state for a real lattice system and then take  it into the Fourier space to study the isotropic superconducting pairing mechanism in the presence of the active Coulomb correlation. Our
approach to define the correlated fermion pairing state $i.e.$ correlated BCS (CBCS) state is-(i) first, construct the Fermi sea state for correlated fermions $i.e.$ correlated Fermi sea(CFS) ground state; (ii) then project out the paired particle states for BCS pairing.
Hence, the CBCS state is defined as
\begin{equation}
\Psi_{CBCS} = P_{BCS}\otimes P_{G} \psi_{FS}
\end{equation}
where, the $P_{BCS}$ and $P_
{G}$ are the BCS pairing operator and the Gutzwiller projection operator respectively. The operator $P_{BCS}$ is acting on the Gutzwiller projected out state from the FS ground state($\psi_{FS}$). The CFS state is defined by making use of the Gutzwiller projection operator as\\
\begin{equation}
\Psi_{CFS} = P_{G}\psi_{FS}
\end{equation} 
where, the Gutzwiller projection operator is defined as \cite{vollahardt1988,anderson1987,himeda-ogata2003,ranjan2003} 
\begin{equation}
P_{G} = \prod_{l} (1 - \alpha n_{l\uparrow} n_{l\downarrow} )
\end{equation}
where, $l$ is the site index and $n_{l\sigma}$ is the occupation number operator at that site with spin states $\sigma = \left\lbrace \uparrow,\downarrow \right\rbrace $. 
The projection operator $P_G$ is preventing the electrons to sit on the same site and this double occupancy is variationally prevented through a parameter $\alpha$ (where, $0\leqslant \alpha \leqslant 1$).\\ 
The FS ground state is defined as
\begin{eqnarray}
\psi_{FS} = \prod_{k} ^ {k_{F}} \sum_{i,j}  e^{i(r_{i}-r_{j}).k}c_{i,\sigma}^{+}c_{j,-\sigma}^{+} \mid 0\rangle \nonumber \\
= \prod_{k} ^ {k_{F}} \sum_{i,j}  e^{i(r_{i}-r_{j}).k} \mid i\sigma, j-\sigma\rangle
\end{eqnarray}
where, $c_{i,\sigma}^{+}(c_{i,\sigma})$ is the fermion creation (annihilation) operator on the i-th site with spin state $ \sigma =   \left\lbrace \uparrow,\downarrow  \right\rbrace  $.  The FS is thus defined by pairwise selection of electrons with opposite momenta and spin
states. Therefore, the CFS state is defined as
\begin{equation}
\Psi_{CFS} = \prod_{l} ( 1 - \alpha n_{l\uparrow} n_{l\downarrow} ) \prod_{k} ^ {k_{F}} \sum_{i,j}  e^{i(r_{i}-r_{j}).k} \mid i\sigma, j-\sigma\rangle
\end{equation}
On operating the BCS pairing operator on the CFS state, the CBCS state is defined as  
\begin{widetext}
\begin{equation}
\Psi_{CBCS} = \prod _{k} ^ {\vert k\vert \geqslant  k_{F}}(u_{k} + v_{k}\sum_{f,g}e^{i(r_{f}-r_{g}).k}c_{f,\uparrow}^{+}c_{g,\downarrow}^{+}) \prod_{l} ( 1 - \alpha n_{l\uparrow} n_{l\downarrow} ) \prod_{k^{'}} ^ {k_{F}} \sum_{i,j}  e^{i(r_{i}-r_{j}).k^{'}} \mid i\sigma, j-\sigma\rangle
\end{equation}
\end{widetext}
where, $u_k$ and $v_k$ are the Bogoliubov coefficients and they follow the conditions for every momentum state \textbf{k}, as
\begin{equation}
|u_{k}|^{2} + |v_{k}|^{2}  = 1
\end{equation}
The CBCS state is not normalized yet, but the normalized state will be taken into account for all the analytical consequences. Meanwhile, an appendix (\textit{Appendix-A}) with a calculation for the normalization constant is  added with this manuscript at the end, as stated before.
\subsection{\label{sec:level2}Model Hamiltonian}
 In one of our previous works on the interplaying aspects of pairing correlation and the Coulomb correlation in a phonon mediated superconductor, we have made use of the reduced zero temperature BCS Hamiltonian\cite{schriefferb,km}. However, we aim to study a more  realistic electron system by setting up a variational state as defined in Eq.(6). In this scenario, it is obvious to define the corresponding Hamiltonian in real space by going over to the Fourier space. Therefore, the real space realization of zero temperature reduced BCS Hamiltonian on a simple cubic lattice for illustration, can now be written as 
\begin{equation}
H = \sum_{i,j} 2\varepsilon_{ij} b_{i}^{+}b_{j}  + \sum_{i^{'},j^{'}}V_{i^{'}j^{'}} b_{i^{'}}^{+}b_{j^{'}}
\end{equation} 
where, $\varepsilon_{ij} = \varepsilon_{0} -2t(cosk_{x}a + cosk_{y}a +cosk_{z}a)$ is the single particle tight binding(TB) energy dispersion for 3D simple cubic lattice, considering nearest neighbour hopping only\cite{kaxiras}. Here, $\varepsilon_{0}$, $t$ and  $a$ are the site energy, nearest neighbour hopping amplitude and the lattice parameter respectively. The pair operators, 
$b_{i}^{+}$ and $b_{i}$ are now in real space written as
  $b_{i}^{+} = \sum_{k} e^{k.r_{i}} b_{k}^{+}$ and $b_{i} = \sum_{k} e^{-k.r_{i}} b_{k}$, where the Cooper pair operators are $b_{k}^{+} = c_{k\uparrow}^{+}c_{-k\downarrow}^{+}$ ; $b_{k} = c_{-k\downarrow}c_{k\uparrow}$.\\
It may be worthwhile remarking here that several superconductors like $Sn-Te, Sn-Sb, Te-Au, Au_{x}Sb_{1-x-y}Te_{y}$ possess simple cubic structures both in the normal and superconducting phases\cite{poon,tsuei,iyo,cohen}.\\ 
The first term within the above Hamiltonian, is representing the paired particle hopping energy, whereas $ V_{i^{'}j^{'}}  =  \sum_{kk^{'}} V_{kk^{'}} e^{-i{(k.r_{i^{'}}-k^{'}.r_{j^{'}})}}$  is the two-body interacting potential. The quantity $V_{kk^{'}}$ is the matrix element used in the BCS formalism to connect two different paired momentum states $k$ and $k^{'}$. 
\\
Beforehand, we do not choose any specific form of the potential; however for our interacting system it is clear that both attractive and the repulsive nature of interactions will appear.
\section{\label{sec:III}Calculation and results}
\subsection{\label{2}Evaluation of ground state energy and Pairing gap functions}
The ground state energy expression can be evaluated by combining Eqs.[6] and [8] as 
\begin{equation}
W =\dfrac{\langle\Psi_{CBCS}|H|\Psi_{CBCS}\rangle}{\langle\Psi_{CBCS}|\Psi_{CBCS}\rangle}
\end{equation}
On evaluating the ground state energy in the CBCS state, we find
\begin{eqnarray}
W = \sum_{k}\frac{ 2\varepsilon_{k}\left( 1-2\alpha+2\alpha^2\right)\vert v_{k}\vert^{2} }{1+2\alpha(\alpha-1)\vert{v}_{k}\vert^{2}} + \nonumber\\
\sum_{k,k^{'}} \frac{P_{k,k^{'}}u_{k^{'}}^{*}u_{k}v_{k^{'}}^{*}v_{k}}{\left[1+2\alpha(\alpha-1)\vert{v}_{k}\vert^{2} \right]\left[1+2\alpha(\alpha-1)\vert{v}_{k^{'}}\vert^{2} \right]   }
\end{eqnarray} 
where, the first term corresponds to the particle-pair energy as measured in the CBCS ground state and the second one represents the interacting matrix elements for the interacting potential. The matrix element $P_{k,k^{'}}$ which  in general connects two different paired momentum states $k$ and $k^{'}$, replaces bare two-body matrix element as prescribed in the usual BCS formalism. The expression for  $P_{k,k^{'}}$ in our case can be cast in the following form:
\begin{equation}
P_{k,k^{'}} = (1+\alpha^{2})\tilde{V}_{k,k^{'}} - 2\alpha \tilde{U}_{k,k^{'}}
\end{equation}
where, $\tilde{V}_{k,k^{'}}$ and $\tilde{U}_{k,k^{'}}$ are the two body interaction terms, appearing with the coefficients which are respectively even and odd functions of $\alpha$ (see \textit{Appendix-B}). The first term in the above equation(Eq.(11)) represents to an
attractive interaction in the pairing mechanism while the second one gives a repulsive interaction.\\
This energy expectation value $W$ is a function of $u_{k}$, ${v}_{k}$  and the 
parameter $\alpha$. We need to minimize the total energy with respect to the Bogoliubov coefficients $u_{k},v_{k}$ as well as with respect to $\alpha$. \\
 The  `superconducting-gap' function is now defined as 
 \begin{equation}
 \Delta^{(\Vec{k})} = \sum_{\langle ij \rangle} V_{i, j}\langle c_{i,\uparrow}^{+}c_{j,\downarrow}^{+}\rangle_{CBCS}
 \end{equation}
where, $V_{ij}$ is the pairing amplitude which connects two lattice sites $i$ and $j$  and for simplicity, we consider the nearest neighbour coupling only. The expectation value in the above equation (Eq.(12)) is evaluated over $\Psi_{CBCS}$ (see \textit{Appendix-C}).\\
On evaluating the paired operator expectation value
and rearranging the odd and even powers of $\alpha$ in the local order parameter, we
define two distinct ‘gaps’ as follows 
\begin{eqnarray}
    \Delta_{1}^{(\Vec{k})} = \sum_{k^{'}}\frac{\tilde{V}_{kk^{'}}u_{k^{'}}v_{k^{'}}(1 + 2\alpha^2)}{1 + 2\alpha(\alpha - 1)\vert v_{k^{'}}\vert^{2}} \\
    \Delta_{2}^{(\Vec{k})} = \sum_{k^{'}}\frac{2\alpha \tilde{U}_{kk^{'}}u_{k^{'}}v_{k^{'}}}{1 + 2\alpha(\alpha - 1)\vert v_{k^{'}}\vert^{2}}
\end{eqnarray}
The physical origin of these two gaps entirely depends on how the double occupancy is being prevented while calculating the pairing amplitude. The even power (here 0 and 2) of $\alpha$  only contributes when the pair operator picks up the electrons pair wise either from the ideal non-interacting FS ground state or from the double occupancy excluded CFS state. However, the contribution of the odd power of $\alpha$ enters in the same calculation while the pair operator acts simultaneously on the ideal FS part as well as the Coulomb correlated FS part. So, that calculation of  odd $\alpha$ part, there is a mixing  between ideal FS and partially double occupancy excluded CFS state.  
The pairing for the first case (see Eq.(13)) is considered to be occurred very close regime to Fermi surface, while the same for the other gap is regarded deep inside the Fermi sea from the surface. In this context, \textit{Appendix-C} is added hereby detailing  the analytical expression of these two gap functions $\Delta_{1}$ and $\Delta_{2}$.\\
However, it is to be noted that, in the zero correlation regime ($i.e.$ $\alpha$ = 0), $\Delta_{2}$ vanishes as expected and $\Delta_{1}$ becomes identical with the definition of ideal BCS gap (for 1-well) (see Eqs.(13) and (14)).\\
\subsection{\label{2}Minimization of ground state energy with respect to both the pairing correlation and Coulomb correlation and self-consistent gap equations}
 The interplaying aspect of the Coulomb correlation and the phonon driven AEE interaction in this pairing mechanism is actively carried out variationally by following extremization with respect to both the Bogoliubov coefficients as well as  the Gutzwiller parameter($\alpha$). For the time being, we take a parametric substitution of the Bogoliubov coefficients $u_{k}$ and $v_{k}$ as 
 \begin{equation}
 u_{k} = sin\theta_{k} ; v_{k} = cos\theta_{k}
 \end{equation}
 On incorporating above transformation, the ground state energy and the `pairing-gaps' become
\begin{eqnarray}
W = \sum_{k}\frac{ 2\varepsilon_{k}\left( 1-2\alpha+2\alpha^2\right)cos^{2}\theta_{k} }{1+2\alpha(\alpha-1)cos^{2}\theta_{k}} + \nonumber\\
\frac{1}{4} \sum_{k,k^{'}} \frac{P_{k,k^{'}}sin2\theta_{k}sin2\theta_{k^{'}}}{\left[1+2\alpha(\alpha-1)cos^{2}\theta_{k} \right]\left[1+2\alpha(\alpha-1)cos^{2}\theta_{k^{'}} \right]   }
\end{eqnarray}
\begin{eqnarray}
\Delta_{1} = \frac{1}{2} \sum_{k^{'}} \frac{\tilde{V}_{kk^{'}}(1+2\alpha^2)sin2\theta_{k^{'}}}{\left[1+2\alpha(\alpha-1)cos^{2}\theta_{k^{'}} \right] }\\
\Delta_{2} = \frac{1}{2} \sum_{k^{'}} \frac{2\alpha \tilde{U}_{kk^{'}}sin2\theta_{k^{'}}}{\left[1+2\alpha(\alpha-1)cos^{2}\theta_{k^{'}} \right] }
\end{eqnarray}
Hence, the total ground state energy $W$ and two gaps $\Delta_{1}$ and $\Delta_{2}$ all three appear as a function of $\theta_{k}$ and $\alpha$.
Therefore, the minimization condition with respect to both the Bogoliubov coefficients and the Gutzwiller parameter runs as follows
\begin{eqnarray}
\frac{\partial W_{k}}{\partial\theta_{k}} = 0  \\
\frac{\partial W_{k}}{\partial\alpha} = 0
\end{eqnarray}
On solving these two simultaneous equations we obtain
\begin{equation}
sin2\theta_{k} = \frac{(A^{'} - 1)\Delta_{k}\varepsilon_{k} + \Delta_{k}\sqrt{\Delta_{k}^{2}(2A^{'} - 1) + A^{'2}\varepsilon_{k}^{2}}}{A^{'}(\Delta_{k}^{2} + \varepsilon_{k}^{2})} 
\end{equation}
where, $A^{'} = 1 - 2\alpha + 2\alpha^{2}$. From the above equation, one can get the analytical expressions of the coefficients $u_{k}$ and $v_{k}$ and making use of them therefore one can obtain the quasi-particle energy. After carrying it out, we find that the quasi-particle energy is consistent with that of the BCS theory at zero correlation. Further, the energy gap at $k = 0$ increases with the rise in $\alpha$ value.\\
Again, utilizing Eq.(21), which is going to set the expressions of the Bogoliubov coefficients $u_{k}$ and $v_{k}$,
 we obtain the self-consistent coupled gap equations for $\Delta_1$ and $\Delta_2$ as
 \begin{widetext}
\begin{eqnarray}
\Delta_{1} = -\frac{1}{2}\sum_{k^{,}}\frac{\tilde{V}_{kk^{'}}(1+2\alpha^2)\Delta_{1}^{'} \sqrt{\Delta_{1}^{'2}(2A^{'}-1)^2 + A^{'2}\varepsilon_{k^{'}}^2}}{A^{'}(2 + A^{'})(\Delta_{1}^{'2} + \varepsilon_{k^{'}}^2)+(A^{'}-1)\sqrt{\Delta_{1}^{'4}(A^{'}-1)^2 + A^{'2}\varepsilon_{k^{'}}^{2}(\Delta_{1}^{'2}+\varepsilon_{k^{'}}^{2})}}\\
\Delta_{2} = -\frac{1}{2}\sum_{k^{,}}\frac{2\alpha \tilde{U}_{kk^{'}}\Delta_{2}^{'} \sqrt{\Delta_{2}^{'2}(2A^{'}-1)^2 + A^{'2}\varepsilon_{k^{'}}^2}}{A^{'}(2 + A^{'})(\Delta_{2}^{'2} + \varepsilon_{k^{'}}^2)+(A^{'}-1)\sqrt{\Delta_{2}^{'4}(A^{'}-1)^2 + A^{'2}\varepsilon_{k^{'}}^{2}(\Delta_{2}^{'2}+\varepsilon_{k^{'}}^{2})}}
\end{eqnarray}
\end{widetext}
These two equations are the  self-consistent gap equations for $\Delta_{1}$ and $\Delta_{2}$ in which $\Delta_{1}(\Delta_{2})$ corresponds to the pairing amplitudes in the momentum state $k$ whereas $\Delta_{1}^{'}(\Delta_{2}^{'})$ is that for the momentum state $k^{'}$. To set up these two equations, an assumption has been taken about the smallness of the amplitude of the gaps in comparison to the single particle energy, $i.e. \frac{\Delta_{i}}{\varepsilon} \ll 1(i = 1,2)$.\\ Meanwhile, the effect of Coulomb correlation is now well established through the Gutzwiller parameter($\alpha$) in above two self-consistent gap equations(Eq.(22) and (23)). Again, in the zero correlation regime, one can not get solution for the $\Delta_{2}$ as expected, whereas the other gap-equation becomes identical with the BCS gap-equation corresponding to the 1-well model. The self-consistent gap equations are now solved in the following section by specifying physically both the range of the pairing potential as well as the range of the momentum states for the pairing fermions. 
\\\\
\subsection{Evaluation of the pairing gaps}
We herewith follow the multi square-well prescription as proposed by Ginzburg and Kirzhnits to get the pairing gaps\cite{ginzburg1,ginzburg-kar, ginzburg-kar1}. In their context, the screened Coulomb potential sets the repulsive square-well potential and the attractive pairing interaction originates from various boson exchange mechanisms. The bosonic mediators such as phonon, charge transfer-exciton, plasmon and magnon(or paramagnon) are mainly considered for pairing in parental metallic phase\cite{rcijp,rc-jha}. However, in our present work we here choose phonon as  the bosonic mediator for the attractive pairing interaction. Several recently discovered superconductors belonging to the hydride and super-hydride family are known to be based upon the phonon exchange pairing mechanism, as pointed earlier\cite{dias,shylin,memley,duan, hamlin,errea}.
\subsubsection{Pairing gap within the attractive well:}
The superconducting  gap  functions  as  in Eq.(22) correspond to the correlated fermion pairing which is physically valid for the entire regimes of Coulomb correlation in terms of Gutzwiller parameter; $\alpha$ [$i.e.$ $0\leqslant \alpha \leqslant 1$]. However, this gap equation is identical to the non-interacting fermion pairing gap, as defined in the conventional BCS formalism in the absence of the Coulomb correlation ($i.e.$ $\alpha = 0$). In this pairing mechanism, pairing potential $\tilde{V}_{k,k^{'}}$  plays the pivotal role and based on the nature of the bosonic mediator, we can set the interaction frequency regime of the attractive potential well. For the time being, we wish to study the phonon-mediated pairing and we set the upper cut-off frequency as the Debye frequency in the normal metallic phase $i.e.$ $\omega_{c}$. Hence, the potential well is defined by following the BCS prescription as:
\begin{eqnarray}
\tilde{V}_{kk^{'}} &=& - V \hspace*{0.3cm}\mbox{for}   - \hbar\omega_{c} \leqslant \varepsilon_{k},\varepsilon_{k^{'}}\leqslant  + \hbar\omega_c \nonumber \\
&=& 0  \hspace*{0.9cm}\mbox {otherwise}
\end{eqnarray}
\hspace{12cm}with $V>0$.\\
An important assumption, we took here is that with the case of isotropic pairing (s-wave) on the Fermi surface, $\Delta_{1} \simeq \Delta_{1}^{'}$ with $\hbar \omega_{c} \ll E_{F}$\cite{km}. The pairing gap $\Delta_1$ can now be evaluated by integrating out the right hand side of Eq.(22) for all the possible momentum states close to the Fermi surface. Thus, one arrives at the following equation
\begin{widetext}
\begin{equation}
1 = \frac{(1+\alpha^2)V}{2}\int_{-\hbar \omega_{c}} ^ {+\hbar \omega_{c}} d\varepsilon_{k} \frac{\Omega(\varepsilon_{k})\sqrt{\Delta_{1}^{2}(A^{'}-1)^2 + A^{'2}\varepsilon_{k}^2}}{A^{'}(2 + A^{'})(\Delta_{1}^{2} + \varepsilon_{k}^2)+(A^{'}-1)\sqrt{\Delta_{1}^{4}(A^{'}-1)^2 + A^{'2}\varepsilon_{k}^{2}(\Delta_{1}^{2}+\varepsilon_{k}^{2})}}
\end{equation} 
\end{widetext}
where, $\Omega(\varepsilon_{k})$  is the single spin electronic density of states(DOS) per unit volume(see fig:2) and $A^{'} = 1-2\alpha +2\alpha^{2}$. \\
\begin{figure}
\includegraphics[scale=0.55]{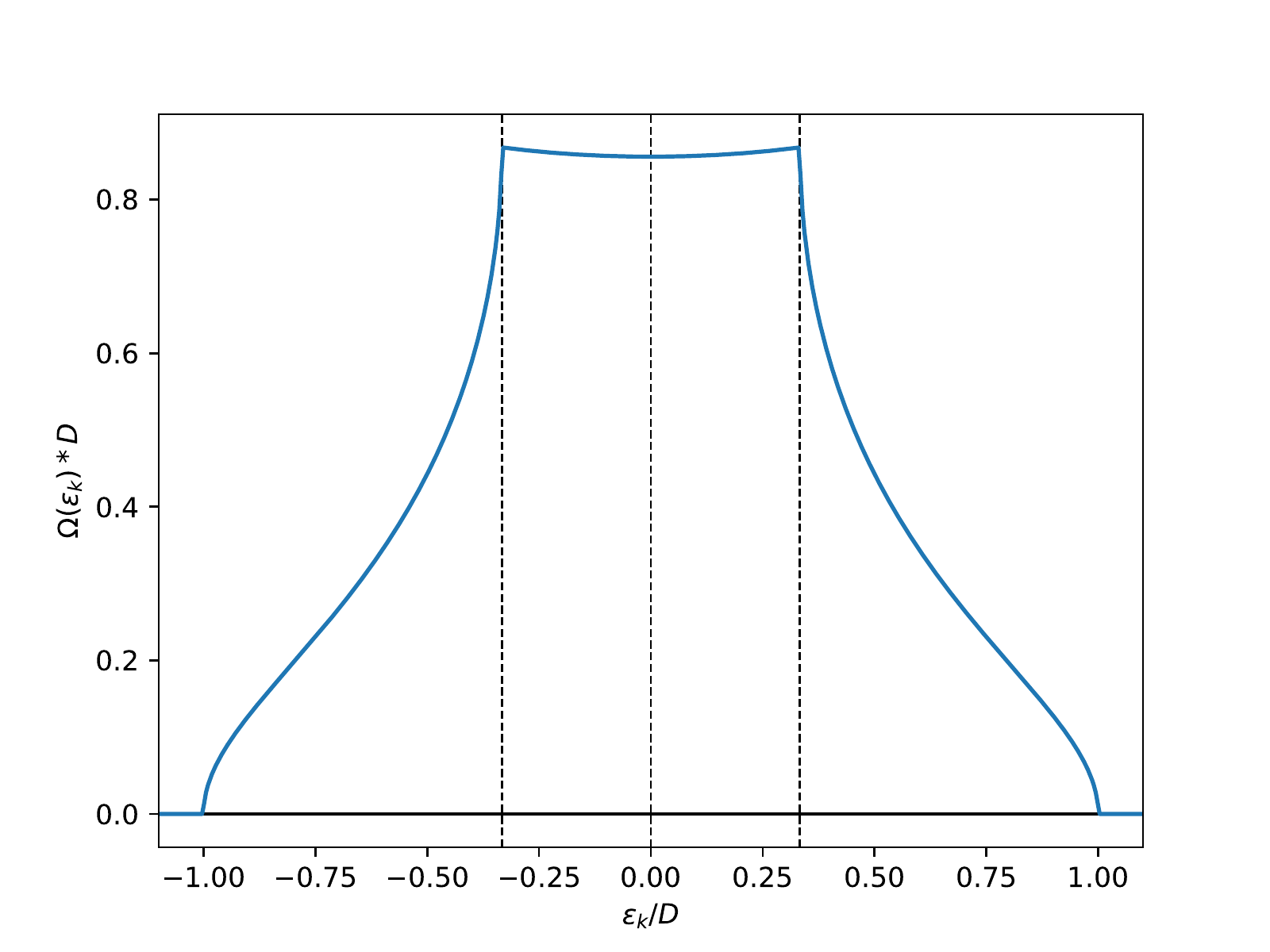}
\caption{\label{fig:wide}The Electronic Density of state(DOS) for simple cubic lattice.}
\end{figure}
Since, our lattice system is a simple cubic(SC) lattice, we need to consider the single spin electron density of state(DOS) to evaluate the integral in Eq.[25]. The DOS for SC lattice from the TB dispersion relation is evaluated by using the elliptic integral of the first kind  as shown in fig:2; where, $D = 6t$ is the half-width of the band with nearest neighbour hopping amplitude $t$ \cite{economou, morita}. Two Van Hove singularity points are there within the band (at $ \pm D/3$) where both real and imaginary part of the Green’s function are continuous but their derivatives show discontinuity\cite{van}. However, within these two singularity points the DOS is almost constant as expected near the Fermi level and hence it is approximated as $\Omega_{0}
$.\\
 The integral on the right hand side of Eq.(25) is now evaluated within these two singularity points such that the range of the attractive well (see Eq.(24)) does not exceed the limit, $\frac{D}{3} \geqslant \hbar\omega_{c}$ . In this context, we take another approximation, that the pairing gap (here, $\Delta_{1}$) is much smaller in amplitude in comparison to the single-particle energy viz.  $\frac{\Delta}{\varepsilon} \ll 1$. Further using this approximations and rearranging the integrand in Eq.(25), we obtain
 \begin{equation}
 \Delta_{1} = 2\hbar \omega_{c} \frac{\left( 1-2\alpha+2\alpha^{2}\right) }{\sqrt{1+4\alpha- 4\alpha^{2}}} exp\left[ -\frac{\left( 1+2\alpha^{2}\right) }{\lambda\left(1-\alpha+2\alpha^{2} \right) }\right] 
 \end{equation}
where, the attractive coupling constant $\lambda$ is the product of $\Omega_0$ and $V$ is defined as the dimensionless  quantity as usual, $i.e.$ $\lambda = V\Omega_0$ . \\

\begin{figure*}
	\centering
	\begin{subfigure}[t]{3in}
		\centering
		\includegraphics[scale=0.6]{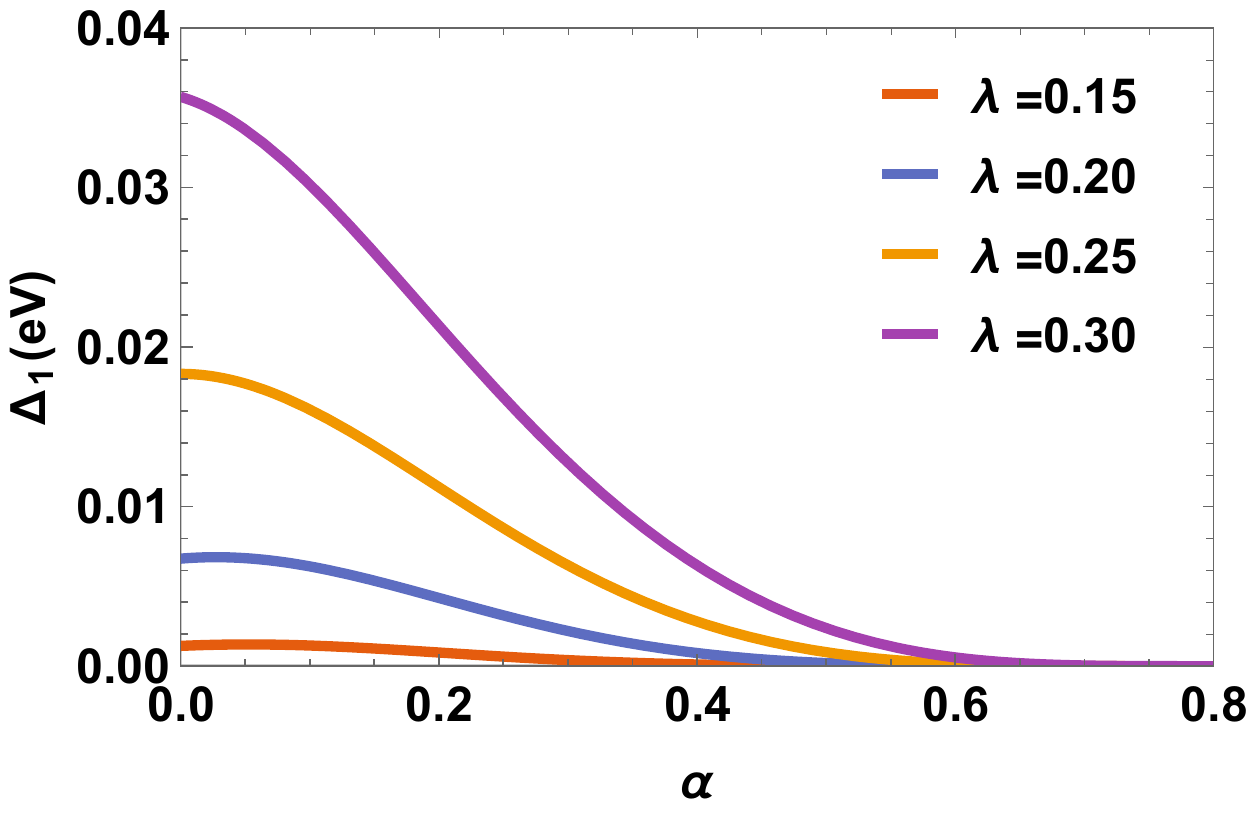}
		\caption{$\Delta_{1}$ vs $\alpha$ for $t = 0.25eV$}	
		  \label{fig:sub1}
	\end{subfigure}
	\quad
	\begin{subfigure}[t]{3in}
		\centering
		\includegraphics[scale=0.6]{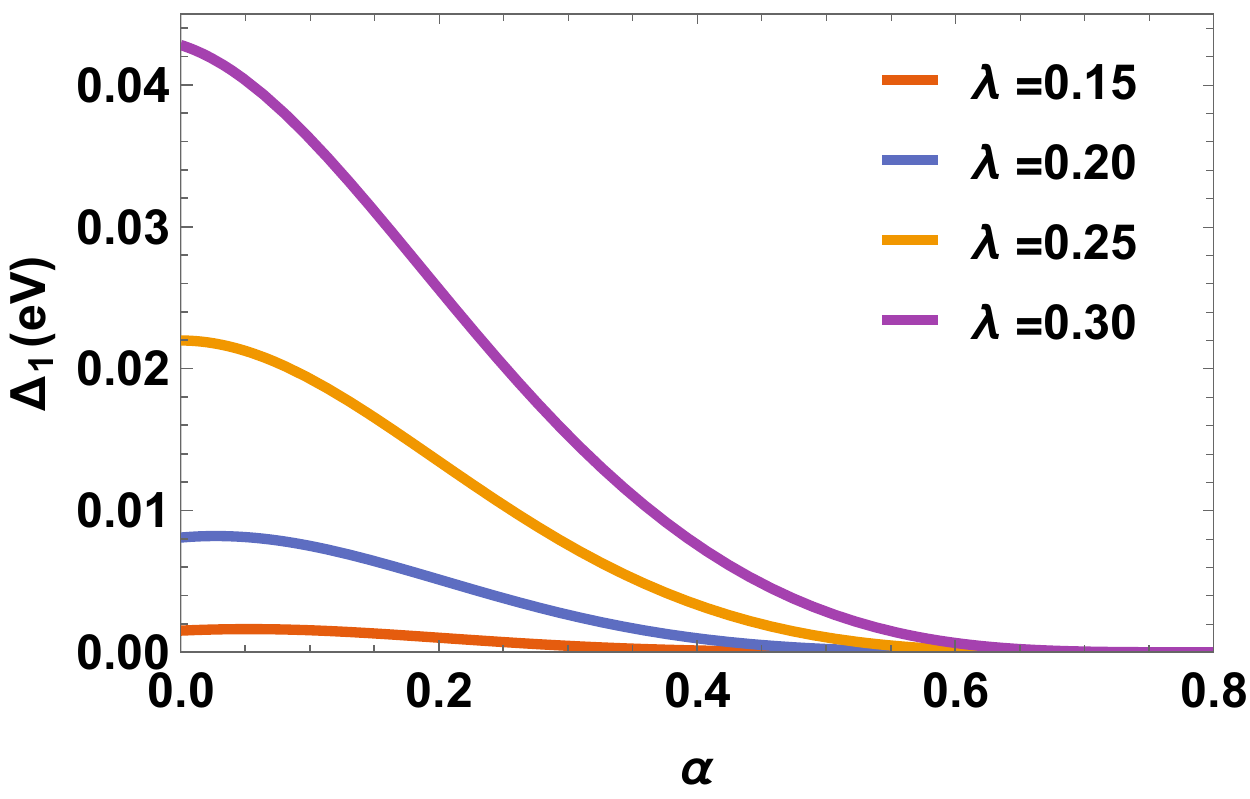}
		\caption{$\Delta_{1}$ vs $\alpha$ for $t = 0.30eV$}
		  \label{fig:sub2}
	\end{subfigure}
	\caption{Variation of $\Delta_{1}$ with $\alpha$ for different coupling values}
	\label{fig:test}
\end{figure*}

The superconducting pairing gap, $\Delta_{1}$ is plotted against $\alpha$ for two different hopping amplitudes $t = 0.25eV$ (fig:3(a)) and $t = 0.30eV$(fig:3(b)) with the range of the attractive potential $\hbar \omega_{c} = 0.05eV$ in the weak coupling limit ($\lambda \leqslant 0.3$). In the zero correlation regime ($\alpha = 0$) it matches exactly with the BCS gap and for a finite coupling constant it decays with $\alpha$. $\Delta_{1}$  however increases with the coupling strength for a given correlation value and the gap closes above a certain correlation value irrespective of the strength of the coupling constant. This gap vanishes for $\alpha = 0.605$ (see fig:3(a) and (b)). Beyond this critical value of $\alpha$, the pairing correlation is no more possible within the attractive square-well and the system
goes towards the Coulomb correlated normal state.\\
\begin{figure*}
	\centering
	\begin{subfigure}[t]{3in}
		\centering
		\includegraphics[scale=0.6]{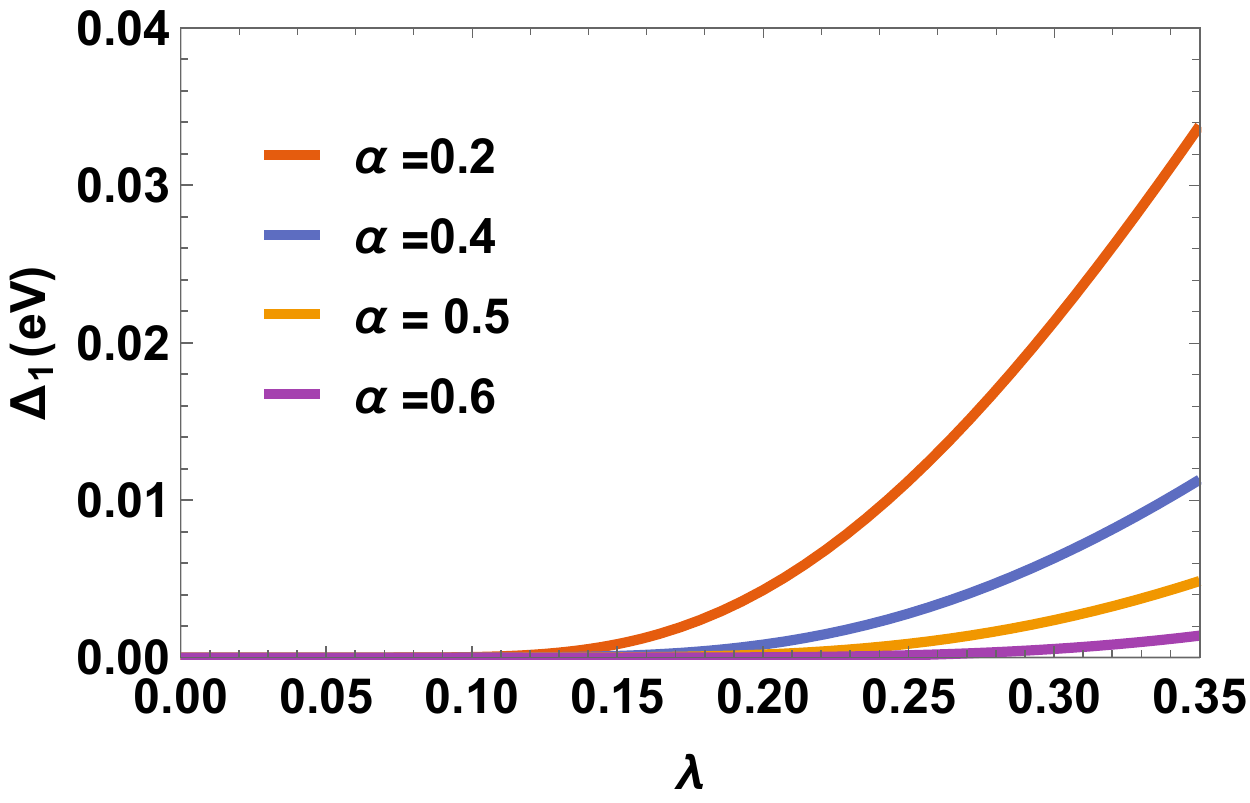}
		\caption{$\Delta_{1}$ vs $\lambda$ for $t = 0.25eV$}	
		  \label{fig:sub1}
	\end{subfigure}
	\quad
	\begin{subfigure}[t]{3in}
		\centering
		\includegraphics[scale=0.6]{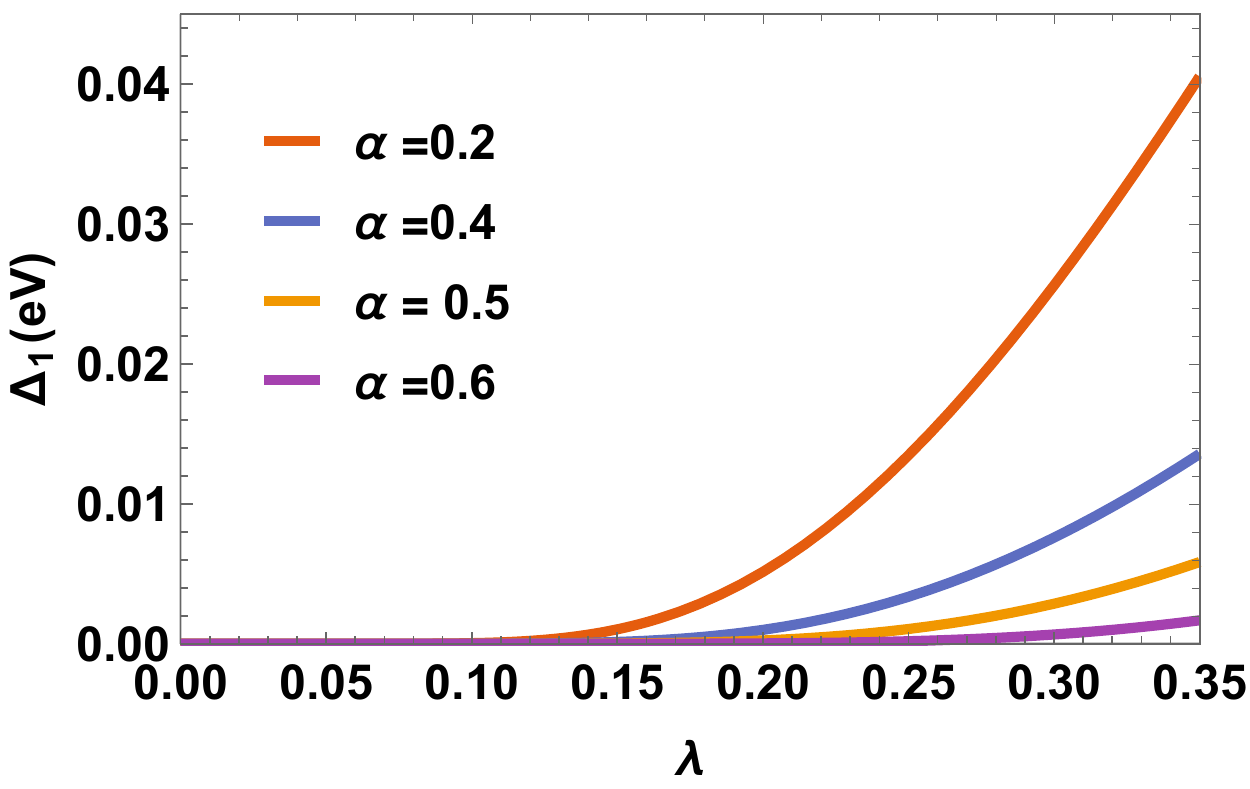}
		\caption{$\Delta_{1}$ vs $\lambda$ for $t = 0.30eV$}
		  \label{fig:sub2}
	\end{subfigure}
	\caption{Variation of $\Delta_{1}$ with $\lambda$ for different coupling values}
	\label{fig:test}
\end{figure*}

In fig: 4, the variation of pairing gap $\Delta_{1}$ with the coupling constant is shown for $t = 0.25eV$ and $t=0.30eV$ (see fig: 4(a) and (b)). The color lines indicate different $\alpha$ values. It is however observed that $\Delta_{1}$ increases with the coupling strength for a given $\alpha$ value but the gap opens up beyond a finite threshold $\lambda$ value for each $\alpha$ value. As expected, $\Delta_{1}$ gets suppressed with an increase in the magnitude of $\alpha$. 
\subsubsection{Pairing gap within the repulsive well:}
The second pairing gap function as in Eq.(23), is somewhat different from the first one in terms of its origin as well as its physical nature and range of pairing potential. At zero correlation ($\alpha = 0$), again this gap function vanishes as predicted(see Eq.(23)) and later we will show that the gap opens up for the finite non-zero correlation value. However, the pairing potential $\tilde{U}_{kk^{'}}$ is repulsive in nature and pairing is taking place in quite an extended regime around the Fermi surface. Hereby, keeping in mind above two criteria, we define the range of the repulsive well as
\begin{eqnarray}
\tilde{U}_{kk^{'}} &=& U \hspace*{0.3cm} \mbox{for}  -\varepsilon_F  \leqslant \varepsilon_{k},\varepsilon_{k^{'}}  \leqslant +\varepsilon_{F}\nonumber\\
&=& 0 \hspace*{0.9cm}\mbox{otherwise}
\end{eqnarray}
\hspace{12cm}with $U>0$.\\
\hspace{9cm}where, $\varepsilon_{F}$ is the Fermi energy.\\
Again, we consider the case of isotropic pairing (s-wave) on the Fermi surface and we assume that $\Delta_{2} \simeq \Delta_{2}^{'}$. The pairing gap $\Delta_{2}$ can now be evaluated by integrating out the right hand side of Eq.(23) for all the possible momentum states and following the form of potential well as in Eq.(27). Hence, one can get the following form
\begin{widetext}
\begin{equation}
1 = \frac{\alpha U}{2}\int_{-\varepsilon_{F}} ^ { +\varepsilon_{F}} d\varepsilon_{k} \frac{\Omega(\varepsilon_{k})\sqrt{\Delta_{2}^{2}(A^{'}-1)^2 + A^{'2}\varepsilon_{k}^2}}{A^{'}(2 + A^{'})(\Delta_{2}^{2} + \varepsilon_{k}^2)+(A^{'}-1)\sqrt{\Delta_{2}^{4}(A^{'}-1)^2 + A^{'2}\varepsilon_{k}^{2}(\Delta_{2}^{2}+\varepsilon_{k}^{2})}}
\end{equation}
\end{widetext}
However, the integration for the $\Delta_{2}$ is not that simple because of its very large frequency regime. However, the DOS for SC lattice shows a flat region around the Fermi level but beyond two singularity points it varies with single particle energy(see fig:2). One can numerically evaluate the integration beyond the singularity points. Meanwhile, the integration is done  over three different regimes of $\varepsilon_{k}$, as: (i) $-\varepsilon_{F}  \leqslant \varepsilon_{k} \leqslant − \hbar \omega_{c} $; (ii) $- \hbar \omega_{c} \leqslant \varepsilon_{k} \leqslant + \hbar \omega_{c} $; and (iii) $ + \hbar \omega_{c} \leqslant \varepsilon_{k} \leqslant  +\varepsilon_{F} $.
 \begin{figure}
  \centering
  \includegraphics[width=1\linewidth]{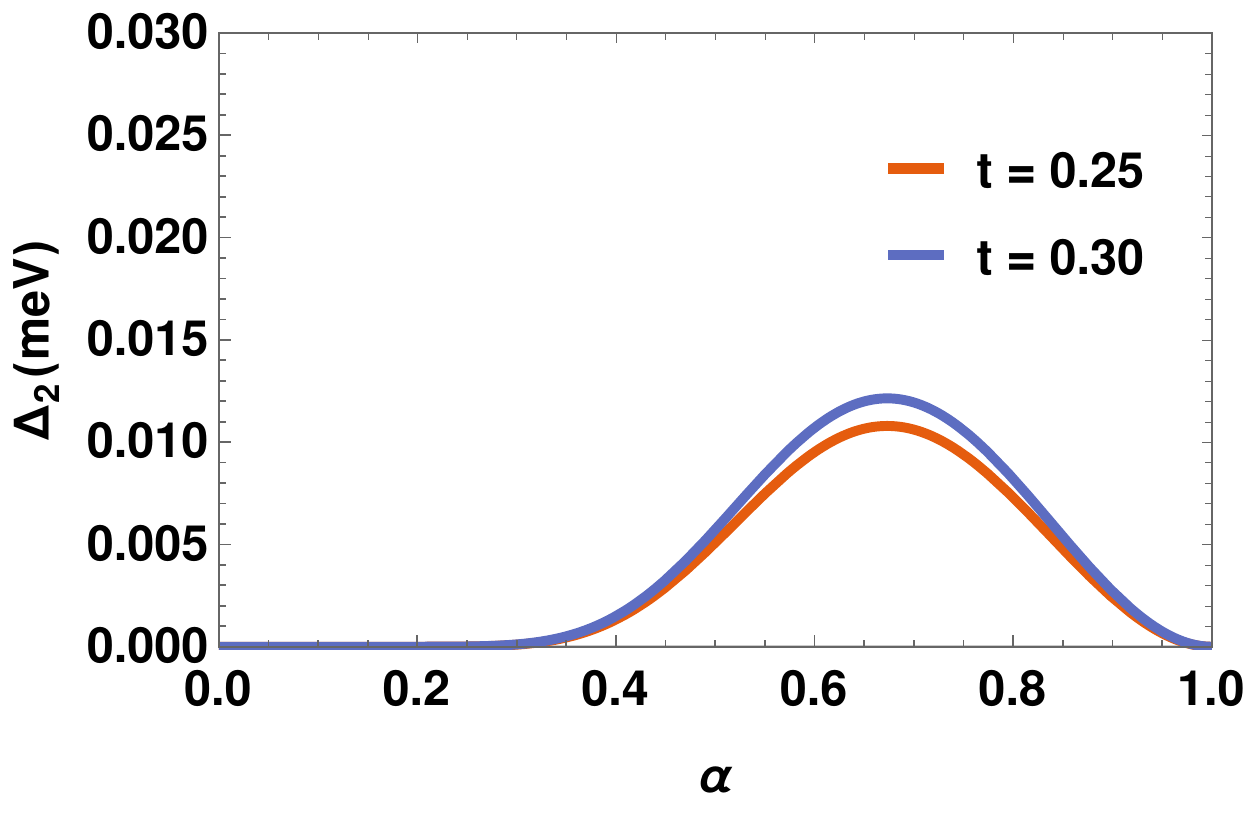}
  \caption{ Variation of $\Delta_{2}$ with $\alpha$}
  \label{fig:sub1}
\end{figure}
\begin{figure*}
	\centering
	\begin{subfigure}[t]{3in}
		\centering
		\includegraphics[scale=0.6]{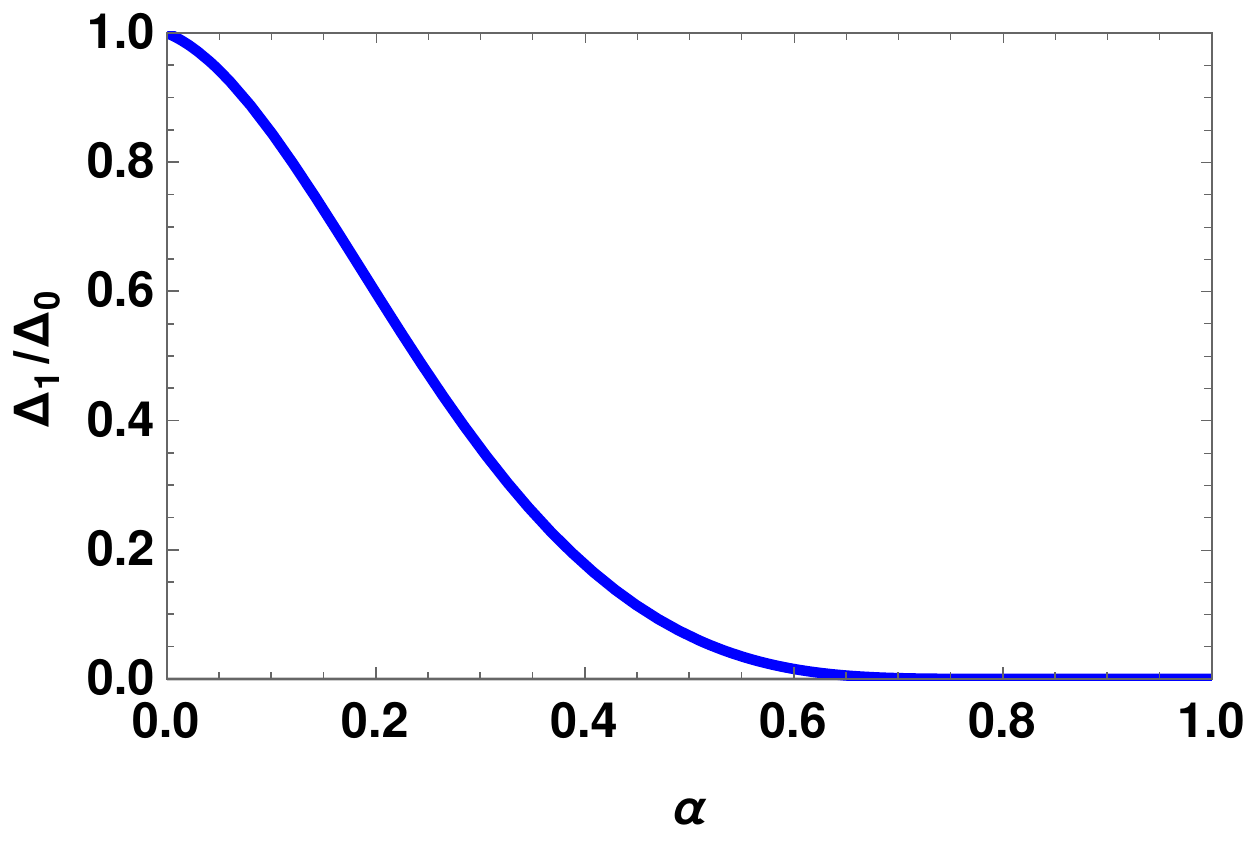}
		\caption{$\frac{\Delta_{1}}{\Delta_{0}}$ vs $\alpha$ for $\lambda = 0.30$}	
		  \label{fig:sub1}
	\end{subfigure}
	\quad
	\begin{subfigure}[t]{3in}
		\centering
		\includegraphics[scale=0.6]{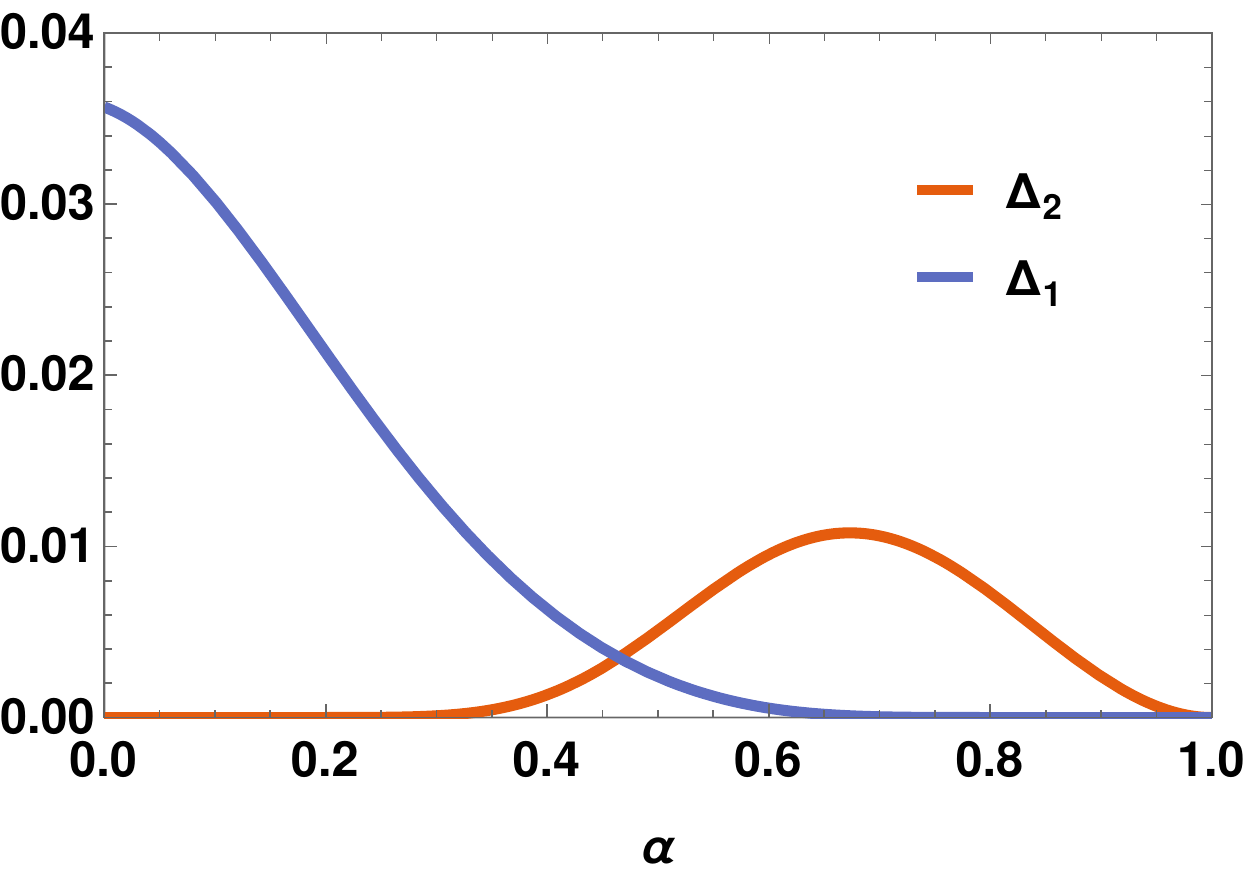}
		\caption{Variation of $\Delta_{1}$ and  $\Delta_{2}$ with $\alpha$ for $t= 0.25$}
		  \label{fig:sub2}
	\end{subfigure}
	\caption{Variation of $\Delta_{1}$ and  $\Delta_{2}$ with $\alpha$ for $t= 0.25$}
	\label{fig:test}
\end{figure*}
Here, in fig:5, variation of $\Delta_{2}$ with $\alpha$ is shown for $t = 0.25eV$ and $t = 0.30 eV$ . In the weak Coulomb correlation regime, $\Delta_{2}$ is found to be zero and it opens up after certain $\alpha$ value. However, $\Delta_{2}$ increases slowly upto certain $\alpha$ value and then again it deceases with large $\alpha$ values. Ultimately, $\Delta_{2}$
vanishes as $\alpha$ approaches very strong correlation regime.\\\\
The above features may be understood as follows:-\\
At lower values of $\alpha$, a sizeable fraction of doubly occupied sites are allowed and therefore Cooper pairs can survive in the 2-well scenario; thereby leading to a finite magnitude of $\Delta_{2}$ with a weightage transfer from $\Delta_{1}$ to $\Delta_{2}$ as well, causing the growth of $\Delta_{2}$ with $\alpha$. However, as $\alpha$ increases further to attain larger values, the doubly occupied sites get mostly eliminated leading to a fall of $\Delta_{2}$.
\section{Conclusion}
\vspace*{0.3cm}
 So far, we have demonstrated how the Coulomb correlation, entering through the Gutzwiller projection scheme of exclusion of the double occupancy on a site, affects the superconducting pairing. However, our two-well model is quite different from our previous work in which we have studied the effect of the Coulomb correlation on paired state\cite{km}. The non-trivial commutation relation among $P_{G}$ and $P_{BCS}$ makes these two superconducting pairing scenarios quite distinct. In other words, we can say that the two alternative routes through which we have introduced the Coulomb correlation in phonon mediated superconducting pairing mechanisms, are itself different from each other in terms of ordering the operators ($i.e.$ $P_{G}$ and $P_{BCS}$) while defining the respective variational state. This leads to two very different consequences in the superconducting phases, as seen here and reported earlier in ref.\cite{km}.\\
For the earlier case, the parental state is a non-interacting ideal FS ground state, and pairing fermions are pair wise picked up from the uncorrelated FS state. The effect of the Coulomb correlation on the phonon mediated superconducting paired state is then studied on the BCS ground state. \\
However, for the current scenario, the FS ground state of the normal phase itself is Coulomb correlated and the boson mediated superconducting pairing is then studied through the BCS pairing operator. The Coulomb correlation is therefore implemented in the parental normal state and the superconducting pairing then takes place in this correlated background.\\
It is to be noted that both of the variational approaches discussed here and in ref.\cite{km} are legitimate under different experimental conditions. In more detail, if one follows the path as in
ref.\cite{km}, the Coulomb correlation on the paired state can be externally adjusted through
the induced pressure or by a sample size-dependent correlation controller\cite{dias,shylin,lado,lado1}.
\\
Some superconducting systems, such as heavy-fermion superconductors, have a strongly
correlated parental normal state for which the superconducting pairing is largely affected
by the strong Coulomb correlation\cite{stelich1979, kuramoto-kitaoka2000}. In view of our present pairing scenario, as discussed
here, one may follow these types of correlated systems for the experimental realization.    
\\
For the present case, we have  variationally treated both the pairing correlation as well as the Coulomb correlation to set up the gap-equations and then the square wells are defined by following Ginzburg-Kirzhnits's prescription\cite{ginzburg1, ginzburg-kar, ginzburg-kar1}. However, the double well structure appears here because of the origin of the pairing scheme as defined in \textit{sec-2}.\\  
In the meantime, $\Delta_{1}$ validates its physical appearance only within the effective attractive well whereas $\Delta_2$ is found within the potential well as in Eq.(27). At $\alpha = 0$, the ratio of  $\frac{\Delta_{1}}{\Delta_{0}}$, (where $\Delta_{0}$ is the ideal BCS pairing gap at zero temperature) is 1 and $\Delta_{2}$ is reduced to zero. In other words, it can be said that if the Coulomb correlation is switched off, the system reduces to the ideal BCS model. For a finite coupling, $\Delta_{1}$ decays with $\alpha$ and beyond $\alpha = 0.605$ it reduces to zero (see fig: 6(a)). This critical $\alpha$ value decreases as the $\lambda$ value is gradually reduced(see fig:3(a) and (b)). However, for a given $\alpha$ value, $\Delta_{1}$ rises in magnitude with the increase in $\lambda$. \\
In summary of our study on the pairing scenario in correlated fermions on the simple cubic lattice,
we have shown the characteristic behaviour of $\Delta_{1}$ and $\Delta_{2}$ with $\alpha$ (see fig: 6(b)). Both the pairing-gaps are found to be consistent throughout the entire physically applicable regime of the Coulomb correlation parameter ($i.e.$ $0 \leqslant \alpha \leqslant 1$). In terms of amplitudes, $\Delta_{2} $ is seen to be much smaller than $\Delta_{1}$.\\
However, all the analytical outcomes reported here are on the basis of FL description of the parental normal phase. The Gutzwiller projection scheme has been implemented to set up CBCS variational state with $\alpha$ as a variational parameter and following the constraint $0\leqslant \alpha \leqslant 1$. In this context, there should be a critical $\alpha$ value beyond which FL description of the parental normal phase itself may not work. For the time being, this limiting criterion is under investigation and in future, we will report it separately.

\begin{acknowledgments}
KM thanks UGC, India for the financial
support [ref. no. 522495(2016)].
\end{acknowledgments}
\appendix
\begin{widetext}
\section{Normalization of CBCS state}
Normalization of the CBCS state is carried out here with some major steps as 
\begin{eqnarray}
\langle \Psi_{CBCS}\vert \Psi_{CBCS} \rangle = \hspace*{10cm} \nonumber \\  \prod_{k^{'}}^{\vert k^{'}\vert\geqslant k_{F}} \prod_{l^{'}}\prod_{k}^{\vert k\vert\geqslant k_{F}} \prod_{l} \langle FS \vert \left( 1-\alpha n_{l^{'}\uparrow} n_{l^{'}\downarrow}\right) \left[ u_{k^{'}} + v_{k^{'}}\sum_{f^{'}g^{'}} e^{ik^{'}.(r_{f^{'}}-r_{g^{'}})} C_{g^{'}\downarrow} C_{f^{'}\uparrow} \right] \nonumber \\ \left[ u_{k} + v_{k}\sum_{fg} e^{-ik.(r_{f}-r_{g})} C_{f\uparrow}^{+} C_{g\downarrow}^{+} \right]\left( 1-\alpha n_{l\uparrow} n_{l\downarrow}\right) \vert  FS \rangle
\end{eqnarray}
The orthogonality criterion of the state eliminates most of the off-diagonal terms and it is obvious that only those terms with equal number of fermion creation and annihilation operators will contribute with a non-zero value to the normalization. Hence, the contributing terms are
\begin{eqnarray}
\langle \Psi_{CBCS}\vert \Psi_{CBCS} \rangle = \hspace{12cm} \nonumber \\ \prod_{k^{'}}^{\vert k^{'}\vert\geqslant k_{F}} \prod_{l^{'}}\prod_{k}^{\vert k\vert\geqslant k_{F}} \prod_{l} [ u_{k^{'}}u_{k} + \sum_{f^{'},g^{'}} \sum_{f,g} v_{k^{'}}v_{k} e^{k^{'}.r_{{f^{'}}g^{'}}} e^{k.r_{fg} } \langle FS \vert C_{g^{'}\downarrow}C_{f^{'}\uparrow} C_{f\uparrow}^{+}C_{g\downarrow}^{+} \vert FS \rangle \nonumber \\ 
-\alpha \sum_{f^{'},g^{'}} \sum_{f,g} v_{k^{'}}v_{k} e^{k^{'}.r_{{f^{'}}g^{'}}} e^{k.r_{fg}} \langle FS \vert n_{l^{'}\uparrow}n_{l^{'}\downarrow} C_{g^{'}\downarrow}C_{f^{'}\uparrow} C_{f\uparrow}^{+}C_{g\downarrow}^{+} \vert FS \rangle - \alpha \sum_{f^{'},g^{'}} \sum_{f,g} v_{k^{'}}v_{k}  e^{k^{'}.r_{{f^{'}}g^{'}}} e^{k.r_{fg}} \nonumber \\  \langle FS \vert  C_{g^{'}\downarrow}C_{f^{'}\uparrow} C_{f\uparrow}^{+}C_{g\downarrow}^{+} n_{l\uparrow}n_{l\downarrow} \vert FS \rangle  + \alpha^{2} \sum_{f^{'},g^{'}} \sum_{f,g} v_{k^{'}}v_{k}  e^{k^{'}.r_{{f^{'}}g^{'}}} e^{k.r_{fg}} \nonumber \\  \langle FS \vert n_{l^{'}\uparrow}n_{l^{'}\downarrow} C_{g^{'}\downarrow}C_{f^{'}\uparrow} C_{f\uparrow}^{+}C_{g\downarrow}^{+} n_{l\uparrow}n_{l\downarrow} \vert FS \rangle ]+ \alpha^{2}  \prod_{k,k^{'}}^{\geqslant k_{F}} \sum_{f^{'},g^{'}} \sum_{f,g} v_{k^{'}}v_{k}  e^{k^{'}.r_{{f^{'}}g^{'}}} e^{k.r_{fg}} \nonumber \\ \langle FS \vert (n_{1\uparrow}n_{1\downarrow} n_{2\uparrow}n_{2\downarrow} + .........+ n_{(N-1)\uparrow}n_{(N-1)\downarrow} n_{N\uparrow}n_{N\downarrow})  C_{g^{'}\downarrow}C_{f^{'}\uparrow} C_{f\uparrow}^{+}C_{g\downarrow}^{+} \vert FS \rangle \hspace{2cm}
\end{eqnarray}
where, $r_{fg} = r_{f} - r_{g}$  and $r_{f^{'} g^{'} } = r_{f^{'}} - r_{g^{'}}$.\\
\begin{eqnarray}
\langle \Psi_{CBCS}\vert \Psi_{CBCS} \rangle =  \prod_{k^{'}}^{\vert k^{'}\vert\geqslant k_{F}} \prod_{l^{'}}\prod_{k}^{\vert k\vert \geqslant k_{F}} \prod_{l} [ u_{k^{'}}u_{k} + \sum_{f^{'},g^{'}} \sum_{f,g} v_{k^{'}}v_{k} e^{ik^{'}.r_{f^{'}g^{'}}} e^{-ik.r_{fg}} \delta_{kk^{'}} \delta_{ff^{'}} \delta_{gg^{'}} \hspace*{2cm}\nonumber \\  -\alpha  \sum_{f^{'},g^{'}} \sum_{f,g} v_{k^{'}}v_{k} e^{ik^{'}.r_{f^{'}g^{'}}} e^{-ik.r_{fg}} \delta_{kk^{'}} \delta_{l^{'}f^{'}} \delta_{l^{'}g^{'}} \delta_{ff^{'}} \delta_{gg^{'}}  -\alpha  \sum_{f^{'},g^{'}} \sum_{f,g} v_{k^{'}}v_{k} e^{ik^{'}.r_{f^{'}g^{'}}} e^{-ik.r_{fg}} \delta_{kk^{'}} \delta_{lf} \delta_{lg} \delta_{ff^{'}} \delta_{gg^{'}}\nonumber \\ + \alpha^{2}  \sum_{f^{'},g^{'}} \sum_{f,g} v_{k^{'}}v_{k} e^{ik^{'}.r_{f^{'}g^{'}}} e^{-ik.r_{fg}} \delta_{kk^{'}} \delta_{l^{'}f^{'}} \delta_{l^{'}g^{'}} \delta_{ff^{'}} \delta_{gg^{'}}] + \alpha^{2} \prod_{k,k^{'}}^{\geqslant k_{F}} \sum_{f^{'},g^{'}} \sum_{f,g} v_{k^{'}}v_{k} e^{ik^{'}.r_{f^{'}g^{'}}} e^{-ik.r_{fg}} \hspace*{2cm}\nonumber \\ 
(\delta_{kk^{'}} \delta_{ff^{'}}\delta_{gg^{'}} + ......) \hspace*{4cm} 
\end{eqnarray}
This calculation is carried out for the momentum state $k = k^{'}$ and using the properties of Kronecker delta function  we arrange the contributing terms as a coefficient of the power of $\alpha$ as :
\begin{eqnarray}
\alpha^{0} &:& \prod_{k} \vert u_{k}\vert^{2} + \vert v_{k}\vert^{2}\nonumber\\ 
\alpha^{1} &:& -2\prod_{k} \vert v_{k}\vert^{2} \nonumber\\ 
\alpha^{2} &:& 2\prod_{k} \vert v_{k}\vert^{2}\nonumber\\ 
\end{eqnarray}
In a compact form this normalization constant 
\begin{equation}
\langle \Psi_{CBCS}\vert \Psi_{CBCS} \rangle = \prod_{k} \left[ 1 -2\alpha(1-\alpha) \vert v_{k}\vert^{2}\right] [\mbox{since,} \hspace {0.2cm}\vert u_{k}\vert^{2} + \vert v_{k}\vert^{2} = 1]
\end{equation}
\section{Ground state energy}
Total ground state energy is calculated as 
\begin{eqnarray}
W = \frac{\langle \Psi_{CBCS} \vert H \vert \Psi_{CBCS}\rangle}{\langle \Psi_{CBCS} \vert \Psi_{CBCS} \rangle}
\end{eqnarray}
The  variational state and the Hamiltonian are as defined in the main text(see Eq.(6) and (8)). However, this is also evaluated in a similar manner as we did for the case of the normalization constant.
 
\begin{eqnarray}
W = \frac{1}{_{CBCS}\langle \Psi \vert \Psi\rangle_{CBCS}}\prod_{k^{'}}^{\vert k\vert\geqslant  k_{F}}\prod_{k}^{\vert k\vert\geqslant k_{F}}\prod_{l^{'}}^{N}\prod_{l}^{N} \langle FS \vert (1-\alpha n_{l^{'}\uparrow} n_{l^{'}\downarrow})[ u_{k^{'}} + v_{k^{'}}\sum_{f^{'}g^{'}} e^{ik^{'}.(r_{f^{'}}-r_{g^{'}})} C_{g^{'}\downarrow} C_{f^{'}\uparrow} ] \nonumber\\
\left( \sum_{i,j} 2\varepsilon_{ij} b_{i}^{+}b_{j}  + \sum_{i^{'},j^{'}}V_{i^{'}j^{'}} b_{i^{'}}^{+}b_{j^{'}} \right)
[ u_{k} + v_{k}\sum_{fg} e^{-ik.(r_{f}-r_{g})} C_{f\uparrow}^{+} C_{g\downarrow}^{+}]\left( 1-\alpha n_{l\uparrow} n_{l\downarrow}\right) \vert  FS \rangle \hspace*{1cm}\nonumber
\end{eqnarray}
\begin{eqnarray}
=\frac{1}{_{CBCS}\langle \Psi \vert \Psi\rangle_{CBCS}} [\prod_{k^{'}}^{\vert k\vert\geqslant k_{F}}\prod_{k}^{\vert k\vert\geqslant k_{F}} \langle FS \vert [1-\alpha(n_{1\uparrow}n_{1\downarrow} +......) +\alpha^{2}(n_{1\uparrow}n_{1\downarrow} n_{2\uparrow}n_{2\downarrow} +....)+...] \hspace*{5cm} \nonumber \\
(u_{k^{'}} + v_{k^{'}}\sum_{f^{'}g^{'}} e^{ik^{'}.(r_{f^{'}}-r_{g^{'}})} C_{g^{'}\downarrow} C_{f^{'}\uparrow}) 
 \left( \sum_{i,j} 2\varepsilon_{ij} b_{i}^{+}b_{j}  + \sum_{i^{'},j^{'}}V_{i^{'}j^{'}} b_{i^{'}}^{+}b_{j^{'}} \right) \hspace*{3cm} \nonumber \\
 ( u_{k} + v_{k}\sum_{fg} e^{-ik.(r_{f}-r_{g})} C_{f\uparrow}^{+} C_{g\downarrow}^{+}) 
(1-\alpha(n_{1\uparrow}n_{1\downarrow} +......) +\alpha^{2}(n_{1\uparrow}n_{1\downarrow} n_{2\uparrow}n_{2\downarrow} +....)+...) \vert FS \rangle ] \hspace*{3cm}\nonumber \\
= \frac{1}{_{CBCS}\langle \Psi \vert \Psi\rangle_{CBCS}} [\prod_{k^{'}}^{ k_{F}}\prod_{k}^{\  k_{F}} [\langle FS \vert (u_{k^{'}} + v_{k^{'}}\sum_{f^{'}g^{'}} e^{ik.r_{f^{'}g^{'}}} C_{g^{'}\downarrow} C_{f^{'}\uparrow})\left( \sum_{i,j} 2\varepsilon_{ij} b_{i}^{+}b_{j}  + \sum_{i^{'},j^{'}}V_{i^{'}j^{'}} b_{i^{'}}^{+}b_{j^{'}} \right)\hspace*{4cm} \nonumber \\
( u_{k} + v_{k}\sum_{fg} e^{-ik.(r_{f}-r_{g})} C_{f\uparrow}^{+} C_{g\downarrow}^{+})\vert FS \rangle] - \alpha [\prod_{k^{'}}^{\vert k\vert \geqslant k_{F}}\prod_{k}^{\vert k\vert \geqslant k_{F}} [\langle FS \vert (n_{1\uparrow}n_{1\downarrow} + n_{2\uparrow}n_{2\downarrow}......) \hspace*{3cm}\nonumber \\
(u_{k^{'}} + v_{k^{'}}\sum_{f^{'}g^{'}} e^{ik.r_{f^{'}g^{'}}} C_{g^{'}\downarrow} C_{f^{'}\uparrow}) \left( \sum_{i,j} 2\varepsilon_{ij} b_{i}^{+}b_{j}  + \sum_{i^{'},j^{'}}V_{i^{'}j^{'}} b_{i^{'}}^{+}b_{j^{'}} \right)( u_{k} + v_{k}\sum_{fg} e^{-ik.(r_{f}-r_{g})} C_{f\uparrow}^{+} C_{g\downarrow}^{+})\vert FS \rangle]+ h.c.]\hspace*{2cm}\nonumber \\
+ \alpha^{2}[\prod_{k^{'}}^{\vert k k_{F}}\prod_{k}^{k_{F}} [\langle FS \vert (n_{1\uparrow}n_{1\downarrow}  n_{2\uparrow}n_{2\downarrow}+......) 
(u_{k^{'}} + v_{k^{'}}\sum_{f^{'}g^{'}} e^{ik.r_{f^{'}g^{'}}} C_{g^{'}\downarrow} C_{f^{'}\uparrow}) 
\left( \sum_{i,j} 2\varepsilon_{ij} b_{i}^{+}b_{j}  + \sum_{i^{'},j^{'}}V_{i^{'}j^{'}} b_{i^{'}}^{+}b_{j^{'}} \right) \hspace*{3cm}\nonumber \\
( u_{k} + v_{k}\sum_{fg} e^{-ik.(r_{f}-r_{g})} C_{f\uparrow}^{+} C_{g\downarrow}^{+})\vert FS \rangle]+ h.c.]]] \hspace{4cm}\nonumber \\
\end{eqnarray}
Herewith, we arrange the coefficients of the power of the Gutzwiller parameter; $\alpha$ as : 
\begin{eqnarray}
\alpha^{0} &:& 2\sum_{k} \frac{\varepsilon_{k}\vert v_{k}\vert^{2}}{ 1 + 2\alpha(\alpha-1)\vert v_{k}\vert^{2}} +  \sum_{kk^{'}} \frac{\tilde{V}_{kk^{'}}u_{k^{'}}u_{k}v_{k^{'}}v_{k}}{ [1 + 2\alpha(\alpha-1)\vert v_{k}\vert^{2}][1 + 2\alpha(\alpha-1)\vert v_{k^{'}}\vert^{2}]}\nonumber\\ 
\alpha^{1} &:& -4\sum_{k} \frac{\varepsilon_{k}\vert v_{k}\vert^{2}}{ 1 + 2\alpha(\alpha-1)\vert v_{k}\vert^{2}} -2 \sum_{kk^{'}} \frac{\tilde{U}_{kk^{'}}u_{k^{'}}u_{k}v_{k^{'}}v_{k}}{ [1 + 2\alpha(\alpha-1)\vert v_{k}\vert^{2}][1 + 2\alpha(\alpha-1)\vert v_{k^{'}}\vert^{2}]}\nonumber\\ 
\alpha^{2} &:& 4\sum_{k} \frac{\varepsilon_{k}\vert v_{k}\vert^{2}}{ 1 + 2\alpha(\alpha-1)\vert v_{k}\vert^{2}} +  \sum_{kk^{'}} \frac{\tilde{V}_{kk^{'}}u_{k^{'}}u_{k}v_{k^{'}}v_{k}}{ [1 + 2\alpha(\alpha-1)\vert v_{k}\vert^{2}][1 + 2\alpha(\alpha-1)\vert v_{k^{'}}\vert^{2}]}\nonumber\\ 
\end{eqnarray}
while evaluating the ground state energy we find that 
$\varepsilon_{k} = \sum_{ij} \varepsilon_{ij}  e^{ik.r_{ij}}$ , with $r_{ij} = r_{i}-r_{j}$; and the connecting matrix elements are
\begin{eqnarray}
 \tilde{V}_{kk^{'}} =  \sum_{i^{'}j^{'}} V_{i^{'}j{'}} e^{i(k^{'}.r_{i^{'}}-k.r_{j^{'}})} \\
\tilde{U}_{kk^{'}} =  - \sum_{i^{'}j^{'}} V_{i^{'}j{'}} e^{i(k^{'}.r_{i^{'}}-k.r_{j^{'}})}
\end{eqnarray}
Here, $\tilde{V}_{kk^{'}}$ and  $\tilde{U}_{kk^{'}}$ are looking similar but they connect the momentum states from two different regimes. We follow the Overhauser formalism to define those regimes\cite{daemen}. For $ \tilde{V}_{kk^{'}}$, the values of $k$ and $k^{'}$ lie within the range $k_{F}<k,k^{'}<k_c$; where, the momentum state $k_{c}$ corresponds to the energy state $\hbar \omega_{c}$. In a similar way, for $\tilde{U}_{kk^{'}}$, the values for both $k$ and $k^{'}$ satisfy $k_{F} <k,k^{'}<k_{max}$; where $k_{max}$ corresponds to an energy state $\sim 2\varepsilon_{F}$.\\
It is interesting to note that if we put $\alpha = 0$, only $\tilde{V}_{kk^{'}}$ is present in that calculation. This is identical to the interacting two body potential as is found in the BCS model. \\ 
The first part of each of the above set of coefficients(Eq.(B3)), signifies the single particle hopping energy. In a similar way, the second part of  these coefficients is the contribution coming from the interacting potential term. 
In the main text, we use a compact expression for the potential energy term as:
\begin{equation}
\sum_{kk^{'}} \frac{[(1+\alpha^{2} \tilde{V}_{kk^{'}}) -2\alpha \tilde{U}_{kk^{'}}]u_{k^{'}}u_{k}v_{k^{'}}v_{k} }{[1 + 2\alpha(\alpha-1)\vert v_{k}\vert^{2}][1 + 2\alpha(\alpha-1)\vert v_{k^{'}}\vert^{2}]} = \sum_{kk^{'}} \frac{P_{kk^{'}} u_{k^{'}}u_{k}v_{k^{'}}v_{k} }{[1 + 2\alpha(\alpha-1)\vert v_{k}\vert^{2}][1 + 2\alpha(\alpha-1)\vert v_{k^{'}}\vert^{2}]}
\end{equation}
\section{Superconducting pairing gaps}
The superconducting-pairing gap function corresponding to the  $CBCS$  state is calculated as follows : 
\begin{equation}
\Delta^{(\Vec{k})} = \sum_{\langle ij \rangle} V_{ij} \langle c_{i\uparrow}^{+} c_{j\downarrow}^{+} \rangle_{CBCS}
\end{equation}
Our system is a simple cubic lattice and for simplicity, we consider the nearest neighbour pairing only.\\
On evaluating the pair-operator expectation value on the CBCS state we have,
\begin{eqnarray}
\Delta = \frac{1}{_{CBCS}\langle \Psi \vert \Psi\rangle_{CBCS}}[\prod_{k^{'}}^{\vert k^{'}\vert\geqslant k_{F}}\prod_{k}^{\vert k\vert \geqslant k_{F}}\prod_{l^{'}}^{N}\prod_{l}^{N} \langle FS \vert (1-\alpha n_{l^{'}\uparrow} n_{l^{'}\downarrow})[ u_{k^{'}} + v_{k^{'}}\sum_{f^{'}g^{'}} e^{ik^{'}.(r_{f^{'}}-r_{g^{'}})} C_{g^{'}\downarrow} C_{f^{'}\uparrow} ] \nonumber\\
\left(\sum_{ij} V_{ij} C^{+}_{i\uparrow} C^{+}_{j\downarrow} \right)
[ u_{k} + v_{k}\sum_{fg} e^{-ik.(r_{f}-r_{g})} C_{f\uparrow}^{+} C_{g\downarrow}^{+}]\left( 1-\alpha n_{l\uparrow} n_{l\downarrow}\right) \vert  FS \rangle] \nonumber\\
= \prod_{k^{'}}^{\vert k^{'}\vert\geqslant k_{F}}\prod_{k}^{\vert k\vert\geqslant k_{F}} \langle FS \vert [1-\alpha(n_{1\uparrow}n_{1\downarrow} +......) +\alpha^{2}(n_{1\uparrow}n_{1\downarrow} n_{2\uparrow}n_{2\downarrow} +....)+...] \nonumber \\
(u_{k^{'}} + v_{k^{'}}\sum_{f^{'}g^{'}} e^{ik^{'}.(r_{f^{'}}-r_{g^{'}})} C_{g^{'}\downarrow} C_{f^{'}\uparrow})\left(\sum_{ij} V_{ij} C^{+}_{i\uparrow} C^{+}_{j\downarrow} \right) ( u_{k} + v_{k}\sum_{fg} e^{-ik.(r_{f}-r_{g})} C_{f\uparrow}^{+} C_{g\downarrow}^{+})\nonumber\\
(1-\alpha(n_{1\uparrow}n_{1\downarrow} +......) +\alpha^{2}(n_{1\uparrow}n_{1\downarrow} n_{2\uparrow}n_{2\downarrow} +....)+...) \vert FS \rangle
\end{eqnarray}
In this calculation, those terms containing upto $\alpha^{2}$ will contribute and the terms with higher powers in $\alpha$ will be eliminated because of the Pauli's exclusion principle. Moreover, on rearranging the contributing terms we find that: \\
For $\alpha^{0}$ term :
\begin{equation}
\langle FS\vert 1. v_{k_{1}^{'}} \sum_{f^{'}g{'}} e^{ik_{1}^{'} .r_{f^{'}g^{'}}} C_{g^{'}\downarrow} C_{f^{'}\uparrow} (\sum_{ij} V_{ij} C^{+}_{i\uparrow}C^{+}_{j\downarrow}) . u_{k_{1}} .1 \vert FS\rangle +....... = \sum_{kk^{'}} \tilde{V}_{kk^{'}} u_{k^{'}}v_{k^{'}}
\end{equation}
for a given momentum state from the ket state $k(\sim k_{F})$, 
\begin{equation}
\tilde{V}_{kk^{'}} = \sum_{ij} V_{ij} e^{i(k^{'}.r_{i}-k.r_{j})} 
\end{equation}
where, the upper cutoff of the $k^{'}$ is  $ k_{c}$; $k_{c}$ corresponds to the maximum accessible momentum state above the Fermi sea in terms of Debye energy of the system(\textit{$i.e.$} $\hbar \omega_{c}$).\\
The phonon mediated superconducting pairing is taken into account through the s-wave channel in the 2-square well scenario. The attractive pairing interaction takes place very close regime to the Fermi surface. Evaluating the pairing amplitude, we find that all the contributions coming from the even power in $\alpha's$ ($i.e.$ 0 and 2) belong to the attractive pairing interaction. Therefore, the product over $k$ and $k^{'}$  for that calculation is running for all those momentum states which are close to the Fermi surface, $i.e.$ $\delta k = \vert k_{F}-k_{c}\vert$. \\
For a given a momentum state $k_{1} = k_{1}^{'} \sim k_{F}$, and $ i = f^{'}$ ; $ j = g^{'}$ one can obtain the first term in the left hand side of the above equation(Eq.(C2)) and later on adding the same for all other momentum states close to the Fermi surface, we get the summed up result as in the right side of the above equation.  \\
In a similar way, the contribution of $\alpha^{2}$ term :
\begin{equation}
 \langle FS\vert (n_{1\uparrow}n_{1\downarrow}n_{2\uparrow}n_{2\downarrow}+...).v_{k_{1}^{'}} \sum_{f^{'}g{'}} e^{ik_{1}^{'} .r_{f^{'}g^{'}}} C_{g^{'}\downarrow} C_{f^{'}\uparrow} (\sum_{ij} V_{ij} C^{+}_{i\uparrow}C^{+}_{j\downarrow}) . u_{k_{1}} .1 \vert FS\rangle +...= \sum_{kk^{'}} \tilde{V}_{kk^{'}} u_{k^{'}}v_{k^{'}}
\end{equation}
The contribution of $\alpha^{1}$ term :\\
This odd $\alpha$ contributing term appears as follows
\begin{equation}
-2\langle FS\vert 1. v_{k_{1}^{'}} \sum_{f^{'}g{'}} e^{ik_{1}^{'} .r_{f^{'}g^{'}}} C_{g^{'}\downarrow} C_{f^{'}\uparrow} (\sum_{ij} V_{ij} C^{+}_{i\uparrow}C^{+}_{j\downarrow}) . u_{k_{1}} .1 \vert CFS\rangle +....... = 2\sum_{kk^{'}} \tilde{U}_{kk^{'}} u_{k^{'}}v_{k^{'}}
\end{equation}
where, for a given momentum state from the ket state $k$,
\begin{equation}
\tilde{U}_{kk^{'}} = -\sum_{ij} V_{ij} e^{i(k^{'}.r_{i}-k.r_{j})}
\end{equation}
 here, the highest accessible momentum state for $k^{'}$ is $k_{max}$, which can correspond to an energy state up to $\sim 2\varepsilon_{F}$.\\
In this case, those momentum states that enter into the pairing amplitude calculation are simultaneously coming from both the ideal non-interacting FS state as well as the CFS state. The CFS state still carries a weightage of allowed doubly occupied sites but to access all those states we have to consider a wide range of $k$ values starting from the deep inside the Fermi sea ($k =0$) to well above the Fermi surface ($\sim 2k_{F}$) as well. Earlier, we had also followed this  criterion to define the repulsive potential well for the $\tilde{U}_{kk^{'}}$ (see Eq.(27) in the main text).  \\
Therefore, the contribution of all three terms are: 
\begin{eqnarray}
\alpha^{0}: \sum_{kk^{'}} \frac{\tilde{V}_{kk^{'}} u_{k^{'}}v_{k^{'}}}{1+2\alpha(\alpha-1)\vert v_{k^{'}}\vert^{2}} \\
\alpha^{1}:  \sum_{kk^{'}} \frac{\tilde{U}_{kk^{'}} u_{k^{'}}v_{k^{'}}}{1+2\alpha(\alpha-1)\vert v_{k^{'}}\vert^{2}} \\
\alpha^{0}: 2\sum_{kk^{'}} \frac{\tilde{V}_{kk^{'}} u_{k^{'}}v_{k^{'}}}{1+2\alpha(\alpha-1)\vert v_{k^{'}}\vert^{2}} 
\end{eqnarray}
Therefore, the two gaps are now defined as 
\begin{eqnarray}
\Delta_1 = \sum_{k^{'}}\frac{\tilde{V}_{kk^{'}}u_{k^{'}}v_{k^{'}}(1 + 2\alpha^2)}{1 + 2\alpha(\alpha - 1)\vert v_{k^{'}}\vert^{2}} \\
\Delta_2 = \sum_{k^{'}}\frac{\alpha \tilde{U}_{kk^{'}}u_{k^{'}}v_{k^{'}}}{1 + 2\alpha(\alpha - 1)\vert v_{k^{'}}\vert^{2}}
\end{eqnarray}


\end{widetext}

\bibliography{apssamp}

\end{document}